\documentclass{rmf-d}
\usepackage{nopageno, rmfbib, multicol,times,epsf,amsmath,amssymb,cite}
\usepackage[latin1]{inputenc}
\usepackage[]{caption2}
\usepackage{graphics}
\usepackage[demo]{graphicx}
\usepackage{hyperref}
\usepackage{lipsum}
\usepackage{wrapfig, blindtext}
\usepackage{comment}
\usepackage{tabularx}
\usepackage{float}
\usepackage{subfigure}
\usepackage{braket}
\usepackage{mathtools}

\def\rmfcornisa{
}

\def\report#1{\vskip0.5\baselineskip\par{\raggedright{\small{\it Report number:} #1}}%
            \vskip0.5\baselineskip\ignorespaces}

\def\rmfcintilla{
}
\clearpage \rmfcaptionstyle 
\begin{document}
\title{  Three-particle scattering amplitudes from lattice QCD
}
\author{ Fernando Romero-L\'opez     }
\address{ Center for Theoretical Physics, Massachusetts Institute of Technology, Cambridge, MA 02139, USA   }
\maketitle
\begin{abstract}
\vspace{1em} Lattice QCD already offers the possibility of extracting three-hadron scattering quantities from first principles. In the last few years, significant progress has been achieved in developing and applying the finite-volume three-body formalism. The formalism is now able to treat physically relevant systems of three mesons, including those with resonances, as well as three-body decays. In this talk, I will review the state of the art, and comment on recent applications to lattice QCD data for systems of three pions and kaons. 
\end{abstract}
\keys{  Lattice QCD, finite volume, three-particle scattering  \vspace{-11pt}}
\report{MIT-CTP/5375 \vspace{-11pt}}
\pacs{   \bf{12.38.Gc, 13.75.Lb}    \vspace{-5pt}}
\begin{multicols}{2}

\section{Introduction}

Understanding the properties of the exotic excitations of the strong force needs first-principles predictions of multi-hadron scattering amplitudes. The formulation of Quantum Chromodynamics (QCD) on the lattice---lattice QCD---promises access to such quantities with systematically improvable uncertainties~\cite{Luscher:1986pf}. Indeed, significant progress has been achieved for low-lying resonances---see Ref.~\cite{Briceno:2017max} for a review. Yet, the exploration of resonances with more complicated decay modes is still at a very preliminary stage. Interesting examples are the Roper resonance, which can decay $N\pi$ and $N\pi\pi$~\cite{Roper:1964zza}, and also other mesonic resonances, such as the $\omega(782)$ and $h_1(1170)$, with predominantly three-pion decay modes~\cite{ParticleDataGroup:2020ssz}.

The foundations of the theoretical framework were laid down by M. L\"uscher for two-particle systems~\cite{Luscher:1986pf,Luscher:1991cf}. The central idea of this formalism is to connect the energy levels extracted from Euclidean correlation functions to infinite-volume scattering quantities. Subsequent theoretical developments~\cite{Rummukainen:1995vs,Lellouch:2000pv,Kim:2005gf,He:2005ey,Bernard:2010fp,Hansen:2012tf,Briceno:2012yi,Briceno:2014oea,Romero-Lopez:2018zyy,Woss:2020cmp,Briceno:2020vgp,Grabowska:2021xkp} achieved a general two-particle formalism that can treat arbitrarily complex two-body systems below the particle production threshold.

More recently, the three-particle problem\footnote{See other recent reviews~\cite{Hansen:2019nir,Rusetsky:2019gyk,Mai:2021lwb}.} in finite-volume has received a lot of attention. In fact, the community has witnessed a blossoming of theoretical developments~\cite{Polejaeva:2012ut,Hansen:2014eka,Hansen:2015zga,Hansen:2015zta,Hansen:2016ync,Briceno:2017tce,Briceno:2018mlh,Briceno:2018aml,Blanton:2019igq,Briceno:2019muc,Jackura:2019bmu,Romero-Lopez:2019qrt,Hansen:2020zhy,Blanton:2020gmf,Blanton:2020jnm,Blanton:2020gha,Hansen:2021ofl,Blanton:2021mih,Blanton:2021eyf,Hammer:2017uqm,Hammer:2017kms,Doring:2018xxx,Romero-Lopez:2018rcb,Pang:2019dfe,Romero-Lopez:2020rdq,Muller:2020wjo,Muller:2020vtt,Mai:2017bge,Guo:2017ism,Klos:2018sen,Guo:2018ibd,Guo:2019hih,Pang:2020pkl,Jackura:2020bsk,Muller:2021uur}, as well as the first applications to lattice QCD data~\cite{Mai:2018djl,Blanton:2019vdk,Mai:2019fba,Culver:2019vvu,Guo:2020kph,Fischer:2020jzp,Alexandru:2020xqf,Hansen:2020otl,Brett:2021wyd,Mai:2021nul,Blanton:2021llb}. The latter has been only possible thanks to technical advances in the extractions of energy levels on the lattice~\cite{HadronSpectrum:2009krc,Morningstar:2011ka,Horz:2019rrn}, with up to hundreds of energy levels available~\cite{Blanton:2021llb}. In fact, some of these studies point towards nonvanishing manifestations of three-particle interactions in finite volume. 

The goal of this talk is to give an overview of the current status of three-particle spectroscopy. I will summarize the theoretical formalism, and describe some of the recent applications to systems of three pions and kaons.

\section{The finite-volume spectrum}

Lattice QCD simulations allow one to stochastically evaluate Euclidean correlation functions:
\begin{equation}
C(t) = \langle \mathcal{O}^\dagger(0) \mathcal{O}(t) \rangle,
\end{equation}
where $\mathcal{O}(t)$ is an operator with some given quantum numbers at Euclidean time $t$. The spectral decomposition of this correlation function reads
\begin{equation}
C(t) = \sum_n |\braket{0|\mathcal{O}|n}|^2 e^{-E_n t},
\end{equation}
where $\{E_n\}$ is the set of energy levels---the spectrum. Therefore, from the time dependence of a correlation function, one can constrain the spectrum and the matrix elements of the operator. In addition, variational techniques enable the determination of several lower lying energy levels~\cite{Luscher:1990ck}.

Since lattice simulations are necessarily performed in a finite box, the measured spectrum corresponds to that of an interacting quantum field theory in a finite volume. In a few cases, the interpretation of these energy levels is simple: if an energy level corresponds to a one-particle state (or stable bound state), these energies are exponentially close to its infinite-volume value~\cite{Luscher:1985dn,Konig:2017krd,Konig:2020lzo}. However, if a state corresponds to a multi-particle state, the connection to infinite-volume is harder to establish. A relevant perspective on this challenge was discussed by Maiani and Testa~\cite{Maiani:1990ca}. In their work, it was shown that one cannot in general obtain on-shell amplitudes from matrix elements of Euclidean correlation functions.

An ingenious strategy to study multi-particle dynamics is to make use of finite-size effects. Restricting systems of particles to a finite volume shifts their energy levels in a way that depends on their interactions~\cite{Huang:1957im}. A very simple example is that of the ground state of two particles of mass $m$. The energy of this state in a box of size $L$ depends asymptotically on the volume as:
\begin{equation}
\Delta E = E-2m = \frac{4\pi a_0}{mL^3} + O(L^{-4}), \label{eq:deltaE}
\end{equation}
where $\Delta E$ is the energy shift, and $a_0$ the $s$-wave scattering length. Therefore, to this order this energy may be mapped into the threshold two-particle amplitude. Similar expansions have been calculated in the literature for more complicated cases. These include higher orders in $1/L$, multi-particle states, the isospin-1 three-pion system, as well as excited states and non-identical particles~\cite{Luscher:1986pf,Detmold:2008fn,Smigielski:2008pa,Hansen:2016fzj,Pang:2019dfe,Romero-Lopez:2020rdq,Grabowska:2021xkp,Muller:2020vtt}.

Perturbative expansions of energy levels can be practical to study some systems of weakly interacting particles, e.g., $\pi^+$ and $K^+$~\cite{Detmold:2008fn,Beane:2007es,Detmold:2011kw,NPLQCD:2020ozd}. They are particularly useful to constrain the two-particle scattering length, as this quantity yields the dominant effect in the finite-volume spectrum. Three-particle effects may also be extracted, although it is technically more challenging. The reason for this is that three-body interactions contribute to the energy shifts with a relative $1/L^3$ suppression with respect to $a_0$. 

\begin{figure}[H]
\centering 
\includegraphics[width=\linewidth]{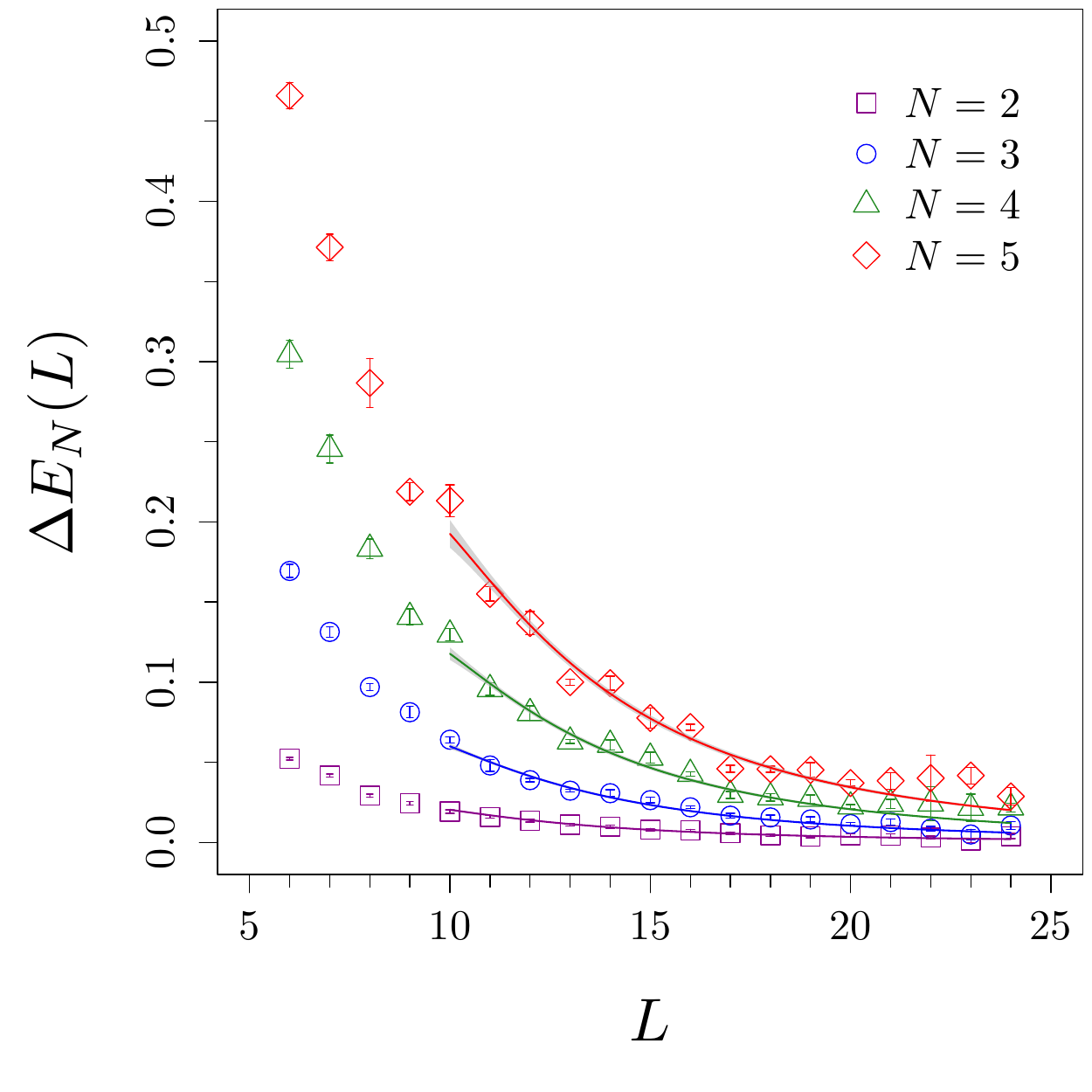}
\vspace{-0.8cm}
\caption{\label{fig:phi4} Energy shift of the $N$-particle ground state as a function of the size of the box, $L$, extracted from lattice simulations in the complex $\varphi^4$ theory~\cite{Romero-Lopez:2020rdq}. A global fit to the expected $L$ dependence yields a statistically significant result for the threshold three-particle amplitude, $\mathcal{\bar{T}} = -19(4) \cdot 10^4$. See Ref.~\cite{Romero-Lopez:2020rdq} for details.}
\end{figure}

A successful example of extracting three-particle quantities in this manner was carried out in Ref.~\cite{Romero-Lopez:2020rdq} using lattice simulations in the complex $\varphi^4$ theory. In that article, the authors fit the volume dependence of the $N$-particle energy levels, with $N=2-5$, to extract the three-particle amplitude at threshold, $\mathcal{\bar{T}}$. This is shown in Fig.~\ref{fig:phi4}, and results in a statistically-significant determination of the amplitude, $\mathcal{\bar{T}} = -19(4) \cdot 10^4$.

\subsection{The L\"uscher formalism}

Unfortunately, properties of resonances cannot be studied with the $1/L$ expansion. Instead, they require the use of a finite-volume formalism that is nonperturbative in $1/L$. For two-particle scattering channels, this is provided by the so-called L\"uscher two-particle quantization condition~\cite{Luscher:1986pf,Kim:2005gf}---abbreviated as QC2. It is an equation whose solutions correspond to the energy levels in finite volume:
\begin{equation}
\det \left[\mathcal K_2(E^*) + F_2^{-1}(E,{\boldsymbol P}, L)\right] \bigg \rvert_{E=E_n} = 0, \label{eq:QC2}
\end{equation}
where $\mathcal K_2$ is the two-particle $K$ matrix evaluated at the center-of-mass (CM) energy, $E^*$, and $F_2^{-1}$ is a known function---the L\"uscher zeta function---that depends on kinematical variables and the box size. The matrix indices of Eq.~(\ref{eq:QC2}) are angular momentum and its third component, $\ell m$. In order to have finite matrices, interactions above some partial wave are neglected. Note that unlike $\mathcal K_2$, $F$ is a finite-volume quantity and it is not diagonal in $\ell m$ space.

\section{The three-particle finite-volume formalism}

The step from two to three particles constitutes an increase in complexity. One reason is that three-particle amplitudes have physical divergences in specific kinematical configurations, which correspond to on-shell propagation of intermediate particles. In addition, three-particle amplitudes depend in general on off-shell two-body interactions, and therefore a separation between two and three-body effects is not well defined. The finite-volume formalism will then require a scheme-dependent intermediate quantity to parametrize three-body interactions, even if all on-shell $S$-matrix elements are uniquely defined.

Early work on the subject showed that the three-body spectrum is determined by $S$-matrix elements~\cite{Polejaeva:2012ut}. Subsequently, the three-body formalism was derived following three different approaches. By chronological order, these are: (i) a generic relativistic effective field theory (RFT)~\cite{Hansen:2014eka,Hansen:2015zga}, (ii) a nonrelativistic effective field theory (NREFT)~\cite{Hammer:2017uqm,Hammer:2017kms}, and (iii) the (relativistic) finite volume unitarity (FVU) approach~\cite{Mai:2017bge,Mai:2018djl}. In its original form, all three approaches dealt with identical (pseudo)scalars with $G$-parity-like symmetry, e.g., a $3\pi^+$ system in the limit of isospin symmetry.

Qualitatively, the RFT approach connects the finite-volume spectrum to the three-particle amplitude by formulating a generic relativistic field theory into finite volume. Note that no specific form of the the theory is required. The NREFT approach uses an effective low-energy nonrelativistic theory. Finally, the FVU approach is derived by means of the unitarity relations of the scattering amplitude in a finite volume.

It is now generally accepted that the three versions should be equivalent. The connection has been explicitly shown for FVU and RFT~\cite{Blanton:2020jnm}. In addition, the equations in the FVU approach and the ``relativized'' formulation of the NREFT~\cite{Muller:2021uur} have the same form.\footnote{Statement based on private communication with A. Rusetsky.} Substantial numerical evidence also supports the equivalence~\cite{Doring:2018xxx,Romero-Lopez:2019qrt}. 

A key point that differs in the different versions is the precise definition of a scheme-dependent intermediate three-particle scattering quantity. In the RFT approach, this object is the three-particle $K$ matrix, $\mathcal K_\text{df,3}$. Here, the subscript ``df'' stands for divergence-free, and indicates that the physical divergences in three-body scattering have been subtracted. Thus, $\mathcal K_\text{df,3}$ is a Lorentz-invariant regular function of the kinematical variables, that has the same symmetries as the underlying theory. These properties will indeed be useful when applying the formalism.

In practice, the three-particle formalism is a two-step process. This is a shared trait of all three versions of the formalism, but for concreteness I focus on the RFT approach. First, the three-particle quantization condition relates the spectrum to the $K$ matrices, $\mathcal K_\text{df,3}$ and $\mathcal K_2$. In the second step, one must solve a set of integral equations that map $\mathcal K_\text{df,3}$ and $\mathcal K_2$ into the three-particle scattering amplitude, $\mathcal M_3$. The latter removes the scheme dependence of $\mathcal K_\text{df,3}$. These steps will be discussed in the next two subsections.

\subsection{The quantization condition}

\begin{figure}[H]
\centering 
\includegraphics[trim={5 35 17.5cm 15},width=\linewidth]{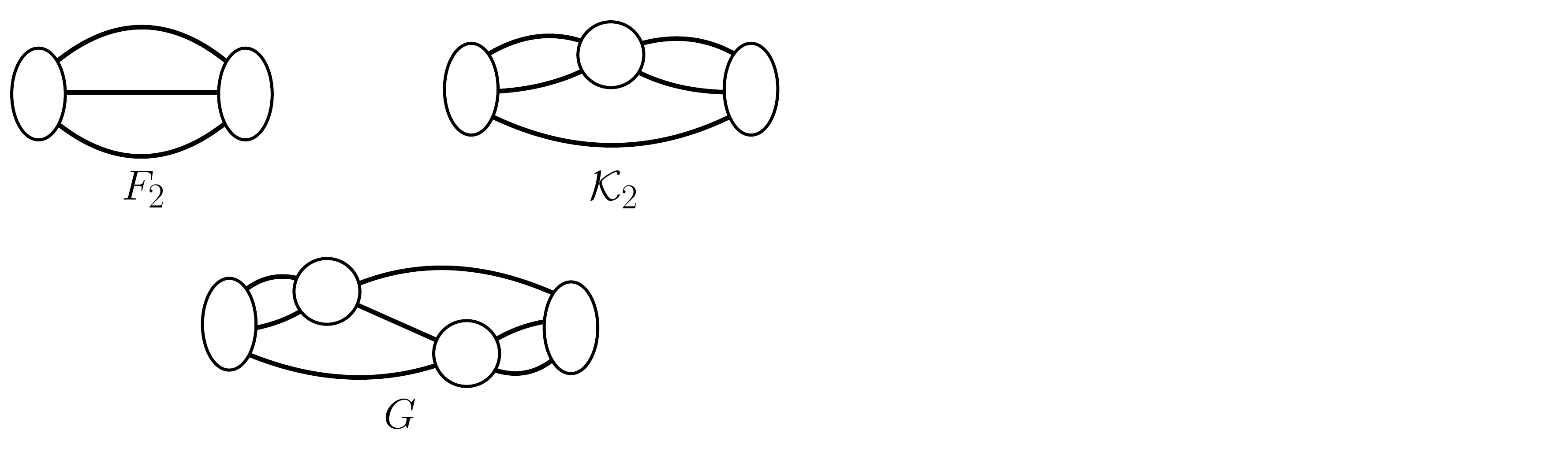}
\caption{Representation of the diagrams that lead to finite-volume effects described by $F_2$, $\mathcal K_2$ and $G$ in Eq.~(\ref{eq:F3}). \label{fig:diags}}
\end{figure}

The three-particle quantization condition (QC3) in the RFT approach for identical scalars reads~\cite{Hansen:2014eka}
\begin{equation}
\det \left[\mathcal K_\text{df,3}(E^*) + F_3^{-1}(E,{\boldsymbol P}, L)\right] \bigg \rvert_{E=E_n} = 0. \label{eq:QC3}
\end{equation}
Although the previous equation looks very similar to Eq.~(\ref{eq:QC2}), there are some fundamental differences. First,  $F_3$ is not a pure kinematical quantity, and it also contains infinite-volume information through the two-particle $K$ matrix, $\mathcal K_2$:
\begin{equation}
F_3 = \frac{F_2}{3} - F_2 \frac{1}{ 1/\mathcal K_2+ F_2 + G} F_2. \label{eq:F3}
\end{equation}
Here, $F_2$ and $\mathcal K_2$ are substantially the same as in the two-body case, and $G$ encodes the finite-volume effects of the one-particle exchange diagrams, in which the spectator particle is switched. The origin of all the elements in Eq.~(\ref{eq:F3}) is depicted schematically in Fig.~\ref{fig:diags}.

In addition, $\mathcal K_\text{df,3}$ and $F_3$ are matrices with indices that characterize three on-shell particles in finite volume. The standard choice is to describe two of the particles---the interacting pair---with angular momentum indices, $\ell m$ in the CM frame. The third particle---the spectator---is described using its finite-volume three momentum, $\boldsymbol k$. A finite dimensionality is ensured by a cutoff function in $\boldsymbol k$ that is built into the formalism, and by neglecting $\ell$ above some value $\ell > \ell_\text{max}$. The latter has to be done in a consistent way to describe two- and three-particle interactions up to the same partial waves~\cite{Blanton:2019igq}.

It is also worth mentioning that several articles by different groups have extended the quantization condition to deal with nonidentical scalar particles~\cite{Hansen:2020zhy,Blanton:2020gmf,Pang:2020pkl,Blanton:2021mih,Mai:2021nul}. Therefore, the formalism is ready to study systems such as the $h_1$ or $\omega$ resonance~\cite{Hansen:2020zhy}, mixed systems of light pseudoscalars, e.g., $\pi\pi K$~\cite{Blanton:2021mih,Blanton:2021eyf}.

\begin{figure}[H]
\centering 
\includegraphics[width=\linewidth]{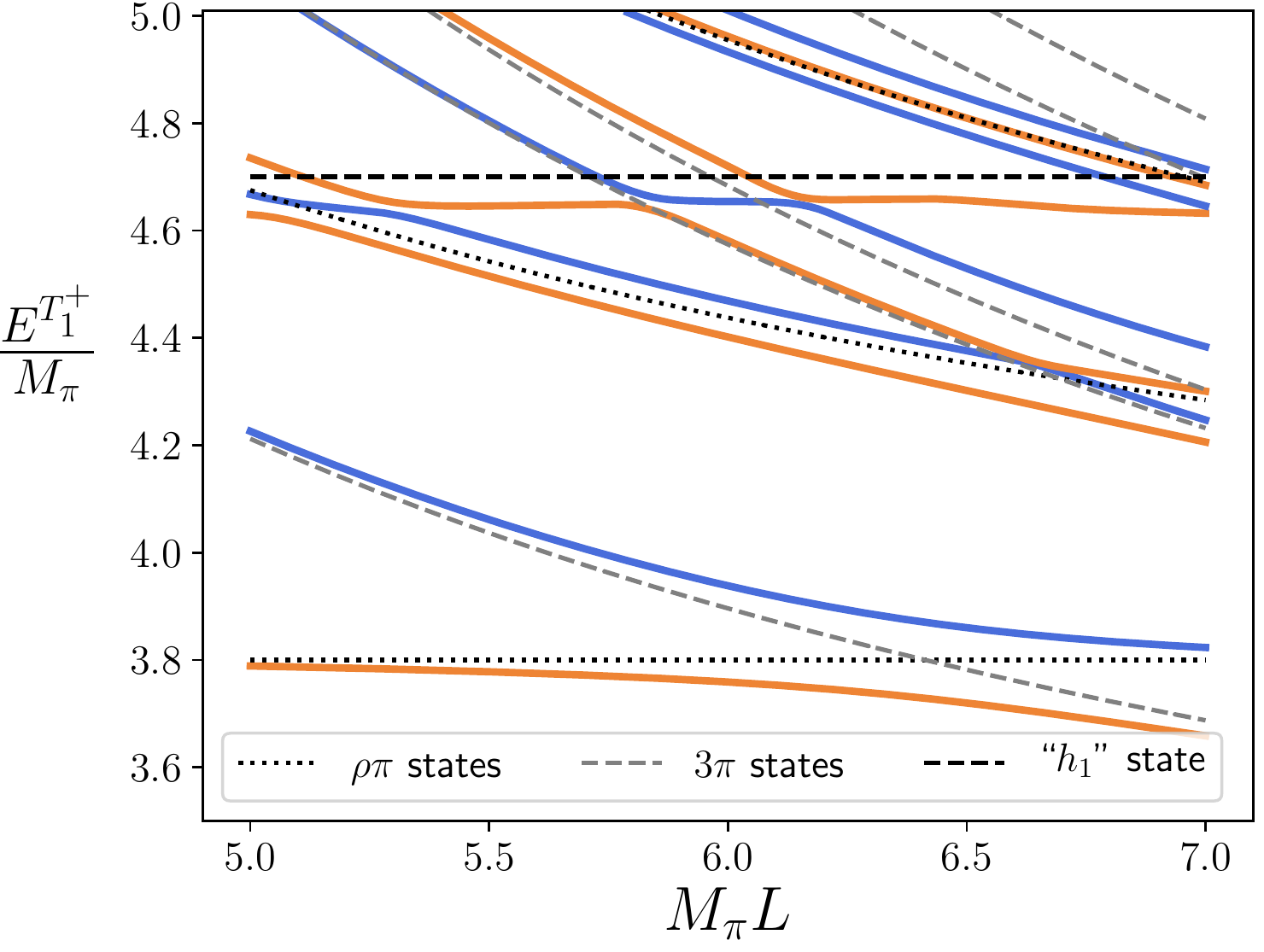}
\vspace{-0.4cm}
\caption{Finite-volume spectrum as a function of $L$ in the $T_1^+$ irreducible representation of the finite-volume symmetry group, that is, a channel that couples to the $h_1$ resonance. It has been generated by solving numerically the three-particle quantization condition for the three-pion isospin-0 channel derived in Ref.~\cite{Hansen:2020zhy}. The two-particle $K$ matrix is chosen such that a resonance similar to the $\rho$ is present. Moreover, the three-particle $K$ matrix contains a pole term to mimic the three-body resonance. Source: Ref.~\cite{Hansen:2020zhy}. \label{fig:isospin}}
\end{figure}

Furthermore, solutions of the quantization condition have been extensively studied in the context of toy models, see e.g., Refs.~\cite{Briceno:2018mlh,Doring:2018xxx}. The idea is to solve the quantization condition for some arbitrary parametrizations of the two- and three-particle interactions. A selected example is Fig.~\ref{fig:isospin}, which shows the volume dependence of the energy levels in a three-pion isospin-0 channel. For this, the three-pion quantization condition derived in Ref.~\cite{Hansen:2020zhy} was used. The two- and three-particle $K$ matrices are chosen such that $\rho$- and $h_1$-like resonances are present.

\subsection{Integral equations}

While $\mathcal{K}_\text{df,3}$ is a very useful quantity to parametrize three-body effects, it is unphysical due to its scheme (or cutoff) dependence. Nevertheless, this can be fixed by solving the integral equations that map the $K$ matrices~\cite{Hansen:2015zga}. Schematically, the procedure is:
\begin{equation}
\mathcal{K}_2, \mathcal{K}_\text{df,3} \xrightarrow[\text{ Integral equations }]{} \mathcal M_3,
\end{equation}
where $\mathcal M_3$ is the full amplitude containing the singularities and analytical structure of three-to-three scattering.

These integral equations have been solved in several works~\cite{Briceno:2018mlh,Jackura:2020bsk,Hansen:2020otl,Mai:2021nul}. In Ref.~\cite{Jackura:2020bsk} different methods are proposed to solve the integral equations when the two-particle subsystem is bound---like a toy model for deuteron-nucleon scattering that neglects spin.

\begin{figure}[H]
\centering 
\vspace{-0.4cm}
\includegraphics[width=\linewidth]{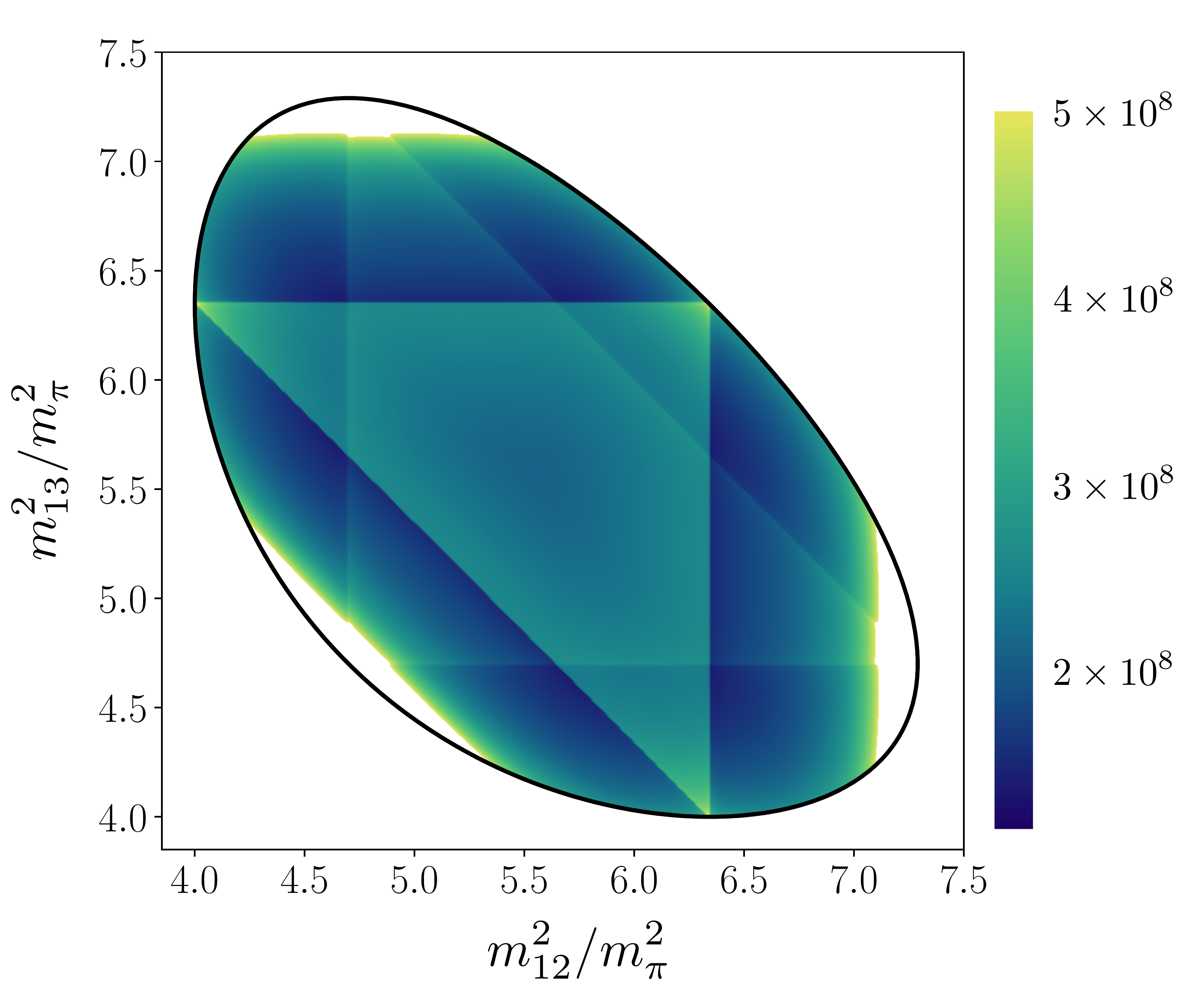}
\vspace{-0.55cm}
\caption{Dalitz plot for three-$\pi^+$ scattering obtained solving the integral equations of the three-body formalism with input from lattice simulations. The color bar indicates the magnitude of $M_\pi^4 |\mathcal M_3|^2$, while the label the incoming squared invariant mass of particles $i$ and $j$ ($m^2_{ij}$). The outgoing kinematics is fixed to be equal to the incoming one. The pion mass is $M_\pi \sim 391$ MeV. More details in the original article: Ref.~\cite{Hansen:2020otl}.\label{fig:dalitz}}
\end{figure}

Another example is that of Fig.~\ref{fig:dalitz}, which shows a Dalitz plot for three-$\pi^+$ scattering obtained after solving the integral equations and using lattice QCD inputs~\cite{Hansen:2020otl}. In this case, the pion mass is $M_\pi \sim 391$ MeV, $\mathcal K_2$ is given by the scattering length measured on the same ensemble, and $\mathcal{K}_\text{df,3}$ is set to zero. The different colors indicate the value of $M_\pi^4 |\mathcal M_3|^2$. Note the presence of divergences in the corners of the plot.

Finally, Ref.~\cite{Mai:2021nul} solves the integral equations in the FVU approach for the three-pion $I=1$ channel, where the $a_0(1260)$ resonance is present. A pole term is included in the three-body intermediate quantity, and the amplitude is analytically continued to the complex plane. Using energy levels from the lattice, new information about the pole positions and branching ratios is deduced.

\subsection{Three-body decays}
Another interesting avenue of the multi-particle formalism in finite volume is the extraction of decays amplitudes from lattice QCD. Due to final-state interactions, decay processes get distorted in a finite volume. In order to obtain the physical amplitude a correction must be applied to the finite-volume matrix element calculated from lattice simulations. In the two-particle sector, this is the so-called Lellouch-L\"uscher formalism~\cite{Lellouch:2000pv} (see also Refs.~\cite{Lin:2001ek,Detmold:2004qn,Kim:2005gf,Christ:2005gi,Meyer:2011um,Hansen:2012tf,Briceno:2012yi,Bernard:2012bi,Agadjanov:2014kha,Briceno:2014uqa,Feng:2014gba,Briceno:2015csa,Briceno:2015tza,Baroni:2018iau,Briceno:2019nns,Briceno:2020xxs,Feng:2020nqj,Hansen:2021usv}). Recently, the formalism for three-particle decays has been worked out in Refs.~\cite{Muller:2020wjo,Hansen:2021ofl} following the NREFT and RFT approaches, respectively. Ref.~\cite{Muller:2020wjo} considers only the case of identical particles, while Ref.~\cite{Hansen:2021ofl} treats systems of three pions in the isospin limit. 

In Ref.~\cite{Hansen:2021ofl}, three hadronic processes for which the formalism can be applied were described. They are (i) the $K \to 3\pi$ weak decay, (ii) the strong isospin-breaking transition $\eta \to 3\pi$, and (iii) the electromagnetic $\gamma^* \to 3\pi$ amplitudes that enter the calculation of the hadronic vacuum polarization contribution to muonic $g-2$. Indeed, one expects that lattice calculations for these can be accessible in the near future, given the recent success in the $K\to\pi\pi$ amplitudes~\cite{RBC:2020kdj}.

Also relevant are the weak decays of $D$ mesons, for which CP violation has been recently confirmed at LHCb~\cite{LHCb:2019hro}. A first-principles prediction will however require the description of four-pion final states in finite volume. 

\section{Results for three-meson scattering}

\begin{figure*}[hbp!]
\vspace{-0.3cm}
\subfigure[Results for $\mathcal K_\text{df,3}^\text{iso,0}$.]{\includegraphics[width=0.5\linewidth]{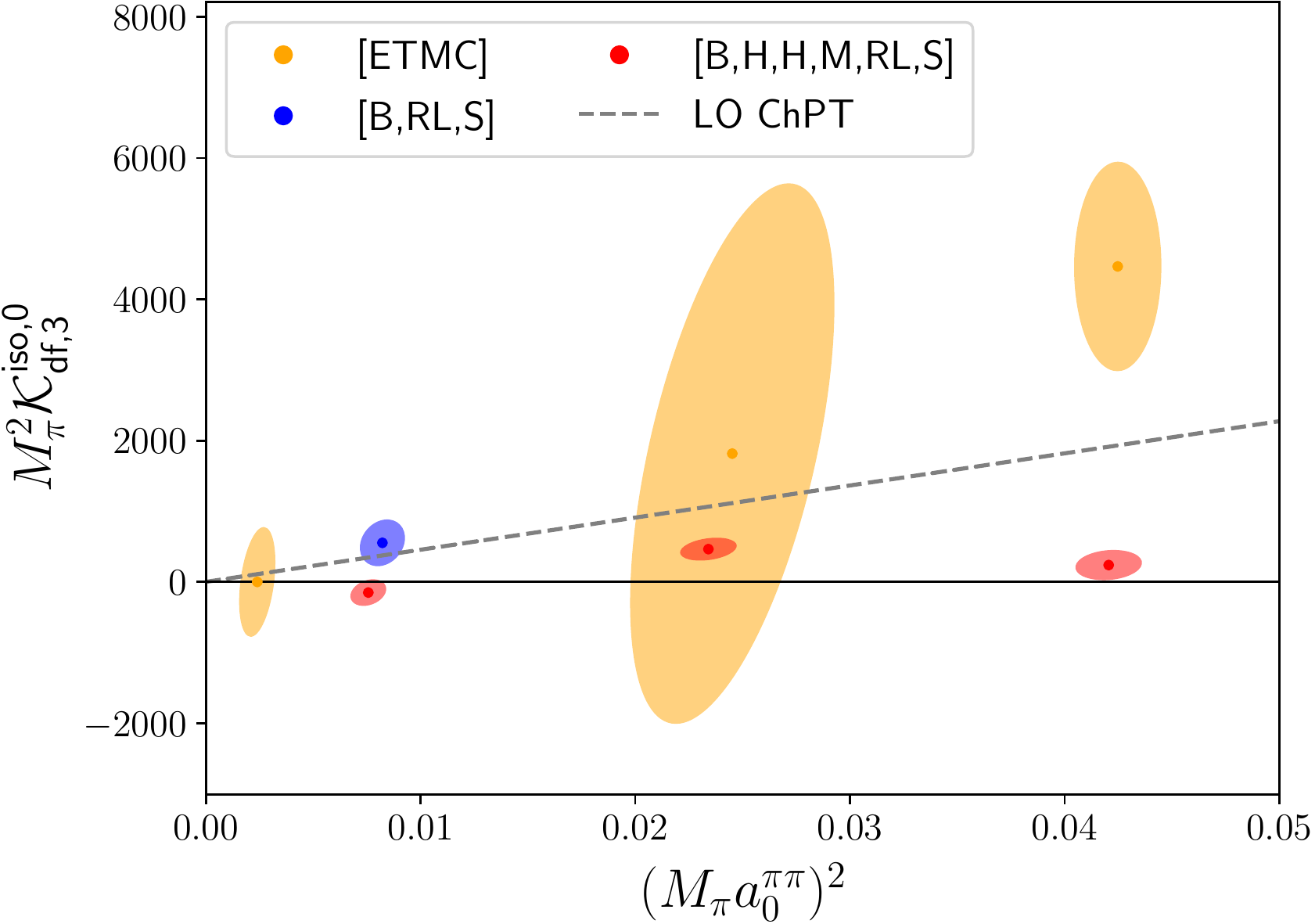}}\label{fig:K0}
\subfigure[Results for $\mathcal K_\text{df,3}^\text{iso,1}$.]{\includegraphics[width=0.5\linewidth]{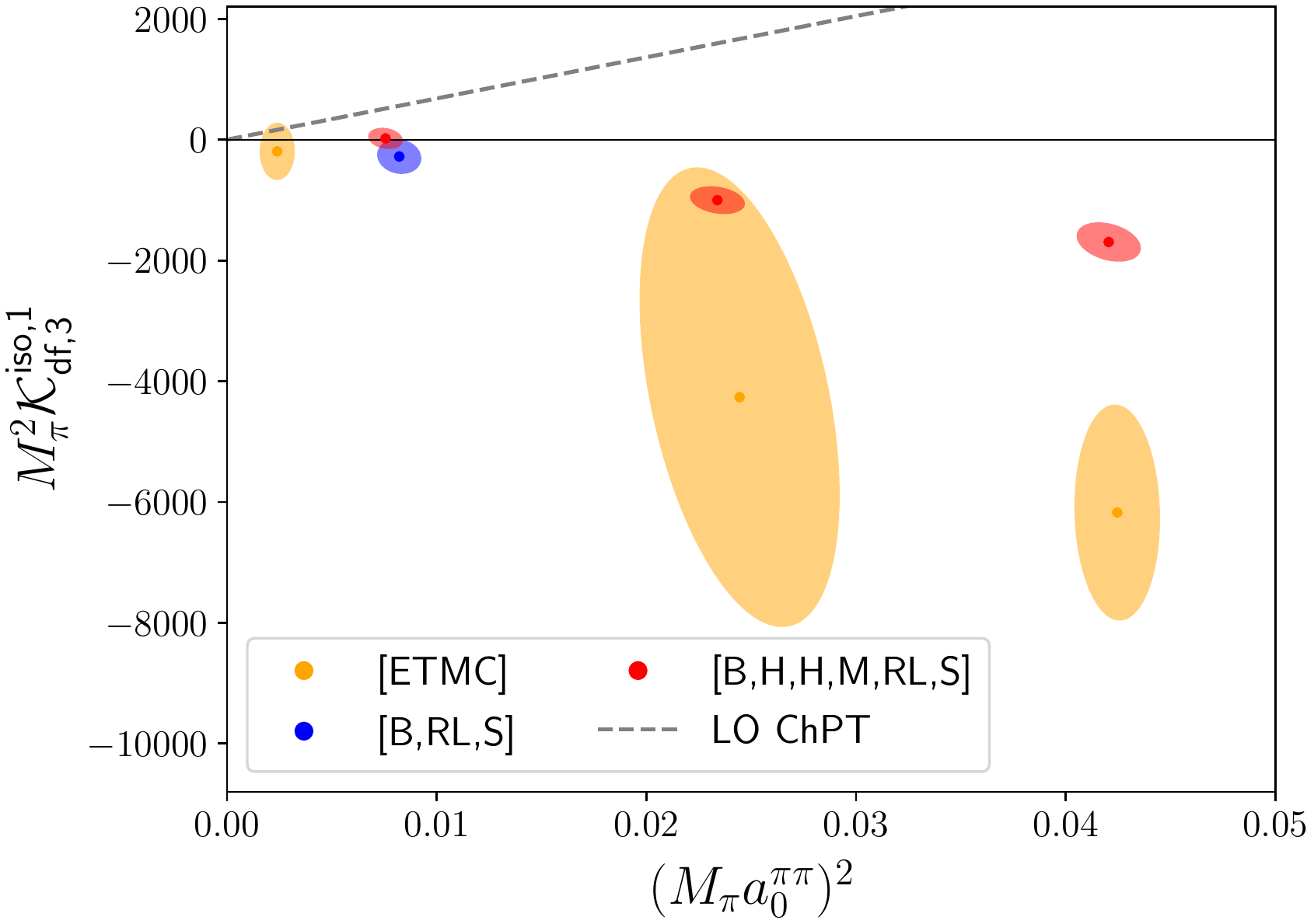}}\label{fig:K1}
\vspace{-0.2cm}
\caption{Results for two terms of the three-particle $K$ matrix for the $3\pi^+$ system in different publications. All results use the RFT formalism, and include only $s$-wave interactions in the two- and three-body sectors. The result in blue ([B,RL,S]) is from Ref.~\cite{Blanton:2019vdk}, orange ellipses ([ETMC]) come from Ref. \cite{Fischer:2020jzp}, and the red ones ([B,H,H,M,RL,S]) from Ref.~\cite{Blanton:2021llb}. \label{fig:Kdfiso}}
\end{figure*}

\begin{figure*}[hbp!]
\centering 
\vspace{-0.4cm}
\includegraphics[width=0.8\linewidth]{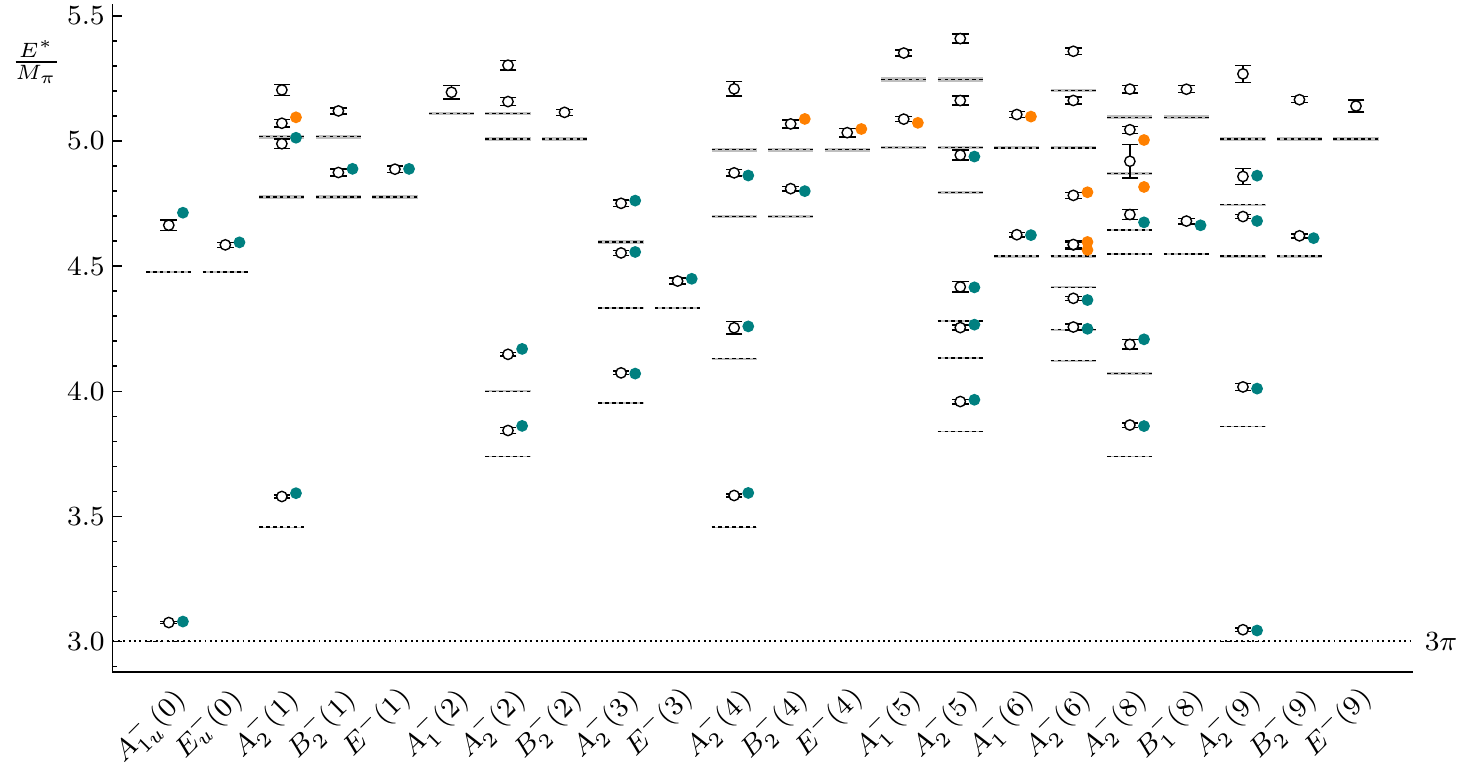}
\vspace{-0.1cm}
\caption{Overview of the finite-volume spectrum of the three positive pions on the N200 CLS ensemble. The vertical axis shows the CM energy, and the various finite-volume irreps and momentum-squared are listed at the bottom. Dashed lines mark the free energy levels, and the open circles denote the interacting energies. Colored circles indicate the central values of the resulting energies from fits to the two- and three-particle quantization condition. Teal circles correspond to energies included in the fits, while orange label those not included. For more details, see Ref.~\cite{Blanton:2021llb}.\label{fig:3pi}}
\end{figure*}

The formalism can already be used to study generic systems of (pseudo)scalar mesons in QCD. But first, it is important to test the methods in controlled setups before turning to more complicated channels. Thus, three pions (or kaons) at maximal isospin constitute an excellent benchmark system for the initial three-body studies.

\subsection{Parametrizing $\mathcal{K}_\text{df,3}$}

In order to constrain the values of $\mathcal{K}_\text{df,3}$, it is essential to find good parametrizations for this quantity. Note that since $\mathcal{K}_\text{df,3}$ is an infinite-volume object, its parametrization will not rely on the finite volume. In Ref.~\cite{Blanton:2019igq}, it was proposed to exploit the symmetry properties of $\mathcal{K}_\text{df,3}$ to carry out a polynomial expansion in the kinematic variables about the three-particle threshold---the threshold expansion of $\mathcal{K}_\text{df,3}$. The expansion parameter will be a set of Mandelstam variables that exactly vanish at the three-particle threshold, for instance, $\Delta \equiv ({s-9 M_\pi^{2}})/{9 M_\pi^{2}}$ and similar quantities.

Let us consider identical particles, such as three $\pi^+$. Imposing that $\mathcal{K}_\text{df,3}$ remains invariant under the symmetries of the theory---particle exchange, parity and time reversal---puts strong constraints on its threshold expansion. In fact, up to quadratic order in $\Delta$, only five terms are present:
\begin{align}
\begin{split}
\mathcal{K}_\text{df,3}&=\mathcal{K}_\text{df,3}^{\mathrm{iso}, 0}+\mathcal{K}_\text{df,3}^{\mathrm{iso}, 1} \Delta+\mathcal{K}_\text{df,3}^{\mathrm{iso}, 2} \Delta^{2}\\&+\mathcal{K}_{A} \Delta_{A}+\mathcal{K}_{B} \Delta_{B} + O(\Delta^3),
\end{split}
\end{align}
where $\mathcal{K}_\text{df,3}^{\mathrm{iso}, 0}, \mathcal{K}_\text{df,3}^{\mathrm{iso}, 1}, \mathcal{K}_\text{df,3}^{\mathrm{iso}, 2}, \mathcal{K}_{A}, \mathcal{K}_{B}$ are real constants, and $\Delta_{A/B}$ are kinematic functions of the Mandelstam variables. The terms that contain ``iso'' in their name (short for ``isotropic'') depend only on the total energy and thus they contribute to $s$-wave interactions. By contrast, $\Delta_{A}$ and $\Delta_{B}$ have an angular dependence: $\Delta_{A}$ corresponds to three particles with $J=0$, but relative $\ell=2$ in the two-particle subsystem, while $\Delta_{B}$ describes an overall $J=2$.

\subsection{Constraining three-body $s$-wave interactions}

The first studies of three-particle interactions have focused on pure $s$-wave interactions~\cite{Mai:2018djl,Blanton:2019vdk,Mai:2019fba,Culver:2019vvu,Guo:2020kph,Fischer:2020jzp,Alexandru:2020xqf,Hansen:2020otl,Brett:2021wyd}. That is, only the $s$-wave phase shift is nonzero, and the three-body parameter is either kept constant, or dependends linearly on $(E^*)^2$. In the RFT approach, the consistent truncation for $s$-wave only interactions is to keep the two leading isotropic terms of $\mathcal{K}_\text{df,3}$:
\begin{equation}
\mathcal{K}_\text{df,3}=\mathcal{K}_\text{df,3}^{\mathrm{iso}, 0}+\mathcal{K}_\text{df,3}^{\mathrm{iso}, 1} \Delta.
\end{equation}

In Fig.~\ref{fig:Kdfiso}, a summary of recent determinations of $\mathcal{K}_\text{df,3}^{\mathrm{iso}, 0}$ and $\mathcal{K}_\text{df,3}^{\mathrm{iso}, 1}$ is shown. In addition, the leading order (LO) chiral perturbation theory (ChPT) prediction~\cite{Blanton:2019vdk} is included for each quantity (dashed line). While several of these results indicate nonzero values for these quantities, there are significant differences between the various works that remain to be understood. Furthermore, the results for $\mathcal{K}_\text{df,3}^{\mathrm{iso}, 1}$ differ significantly from the ChPT prediction, which suggests important effects from higher orders.

Interestingly, at LO in the chiral expansion, $\mathcal{K}_\text{df,3}$ is trivially related to the full amplitude~\cite{Blanton:2019vdk}:
\begin{equation}
\mathcal{M}_3 - \mathcal D^\text{LO}  = \mathcal{K}_\text{df,3} \left[1 + O(M^2/F^2)\right],
\end{equation}
where $\mathcal D^\text{LO}$ is a subtraction term that cancels the divergences of the one-particle exchange diagrams (see Appendix S2 in Ref.~\cite{Blanton:2019vdk}).
In other words, the scheme dependence in this quantity arises at next-to-leading order (NLO) in ChPT. Therefore, a nonvanishing value of $\mathcal{K}_\text{df,3}$ is meaningful despite its scheme dependence. Moreover, note that a NLO prediction could be derived using the results of Ref.~\cite{Bijnens:2021hpq} with the appropriate subtraction scheme.

\subsection{Beyond $s$-wave interactions}

The study of three-body interactions is not limited to the leading $s$-wave effects. This was shown in Ref.~\cite{Blanton:2021llb}, where three-$\pi^+$ and three-$K^+$ interactions were studied including $d$-wave effects. In order to achieve this, it was crucial to determine a large number of energies in each channel---$O(100)$ energy levels per ensemble in different irreps and frames were determined. An example of the measured energy levels is shown in Fig.~\ref{fig:3pi} for $3\pi^+$ on the N200 CLS ensemble ($M_\pi \sim 280$ MeV). The hollow markers indicate the measured spectrum, while colored points are predictions from the QC3. Teal circles denote energies included in the fit, while orange ones are those not included. Note that the fit range for the fits to the quantization condition does not go above the $E^*=5M_\pi$ inelastic threshold, even if the quantization condition still seems to describe those energy levels appropriately. This suggests that inelasticities may not be very relevant. 
 
\begin{figure}[H]
\centering 
\vspace{-0.04cm}
\includegraphics[width=\linewidth]{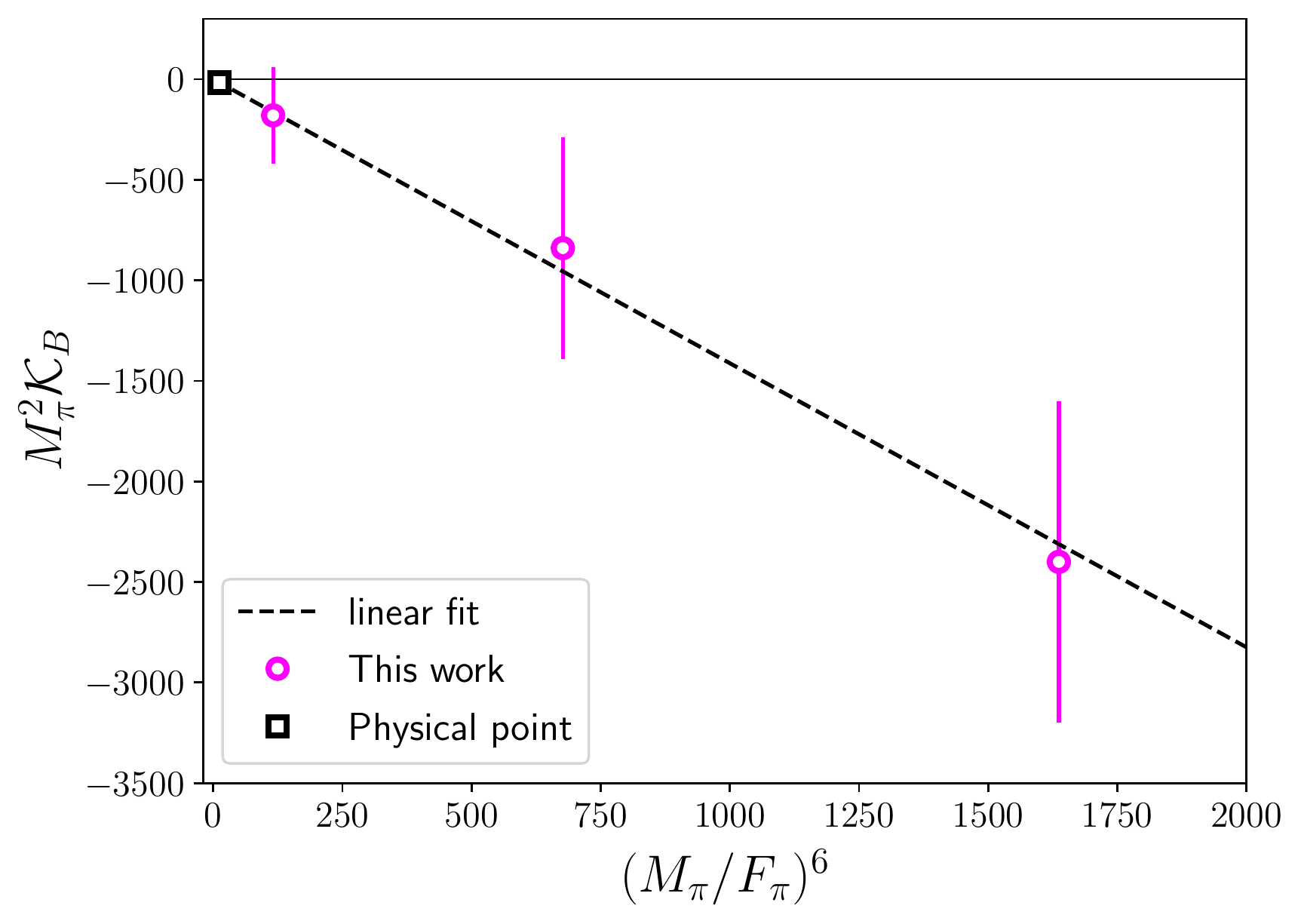}
\vspace{-0.55cm}
\caption{Lattice determination of the $d$-wave three-$\pi^+$ scattering quantity, $\mathcal K_B$. The results for three different values of the pion mass are shown, along with the a fit to the expected chiral behavior, $M_\pi^2\mathcal K_B \propto (M_\pi/ F_\pi)^6$. The physical point is marked with a empty square. Source: Ref.~\cite{Blanton:2021llb}.\label{fig:KB}}
\end{figure}

\begin{figure}[H]
\centering 
\vspace{-0.04cm}
\includegraphics[width=\linewidth]{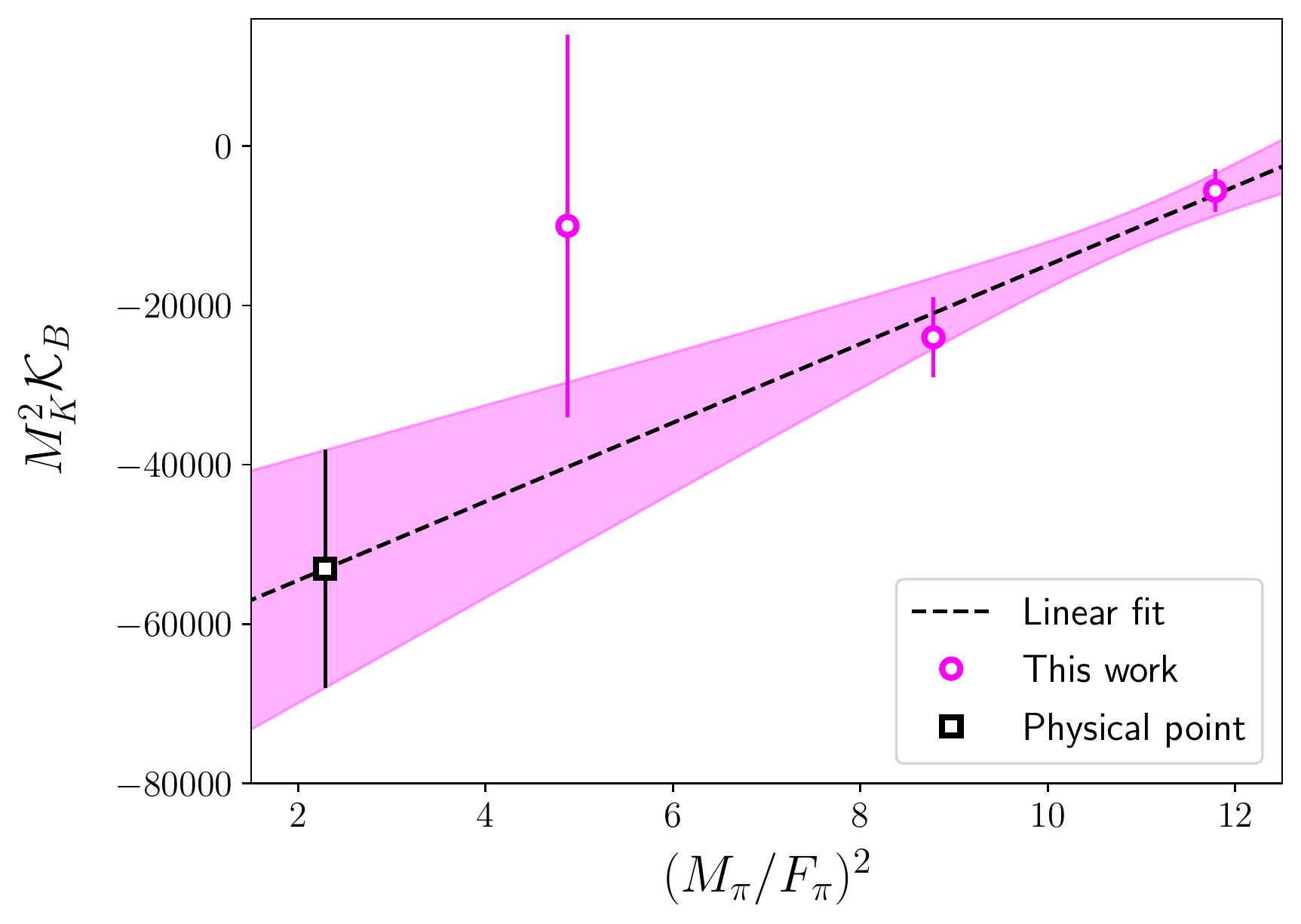}
\vspace{-0.55cm}
\caption{Lattice determination of $\mathcal K_B$ for $3K^+$ scattering. Magenta circles show the results for three different ensembles, and they are plotted against $(M_\pi/F_\pi)^2$. A linear extrapolation to the physical point is also included. Source: Ref.~\cite{Blanton:2021llb}.\label{fig:KB_K}}
\end{figure}

An exciting result of Ref.~\cite{Blanton:2021llb} is the determination of $\mathcal{K}_B$ with statistical significance in most of the ensembles, even if this term is expected to produce a subleading effect in the energy shifts. It turns out that it is the leading contribution of $\mathcal{K}_\text{df,3}$ to the energy levels in nontrivial irreducible representations of the spatial symmetry group. In Fig.~\ref{fig:KB}, results for $\mathcal{K}_B$ for three $\pi^+$ are shown. As can be seen, good agreement with the chiral expectation, $M_\pi^2\mathcal K_B \propto (M_\pi/ F_\pi)^6$, is found. Interestingly, at the physical point this quantity is very small, and it would be very hard to determine in direct simulations with physical pions. Finally, Fig.~\ref{fig:KB_K} shows the same object for kaons, along with an extrapolation to the physical point.

\section{Conclusion and outlook}

Three-particle spectroscopy is progressing rapidly, both in the theoretical developments, as well as applications to QCD. Simple systems of pseudoscalar can be studied, and the first steps towards three-particle resonances have been taken~\cite{Hansen:2020zhy,Mai:2021nul}.

From the theoretical perspective, there are still some open questions. Examples are the formalism for systems of three particles with spin, and multi-channel three-particle processes. It is also to be expected that more applications of the formalism will appear in the following years, involving three-body systems of growing complexity.
 
The long-term aspiration is the first-principles computation of properties of higher-lying resonances, such as $XYZ$ particles and other exotics. This will require further theoretical developments---either in the form of $N$-particle quantization conditions, or alternative approaches~\cite{Bulava:2019kbi,Bruno:2020kyl,Garofalo:2021bzl,Bulava:2021fre}.

\section*{Acknowledgements}

I would like to thank the organizers of the 19th HADRON Conference for the opportunity to give this plenary talk. Special thanks to Akaki Rusetsky, Ben H\"orz and Steve Sharpe for useful comments on this manuscript. 

This work is supported by the U.S. Department of Energy, Office of Science, Office of Nuclear Physics, under grant Contract Numbers DE-SC0011090 and DE-SC0021006.

\end{multicols}
\medline

\begin{multicols}{2}

\end{multicols}
\end{document}